\documentclass[aps,prd,twocolumn,showpacs,groupedaddress,nofootinbib]{revtex4-1}

\usepackage{graphicx}
\usepackage{dcolumn}
\usepackage{bm}
\usepackage{url}
\def\actaa{Acta Astron.}%
\def\apj{ApJ}%
\def\apjl{ApJ}%
\def\apss{Ap\&SS}%
\def\aap{A\&A}%
\def\icarus{Icarus}%
\def\mnras{MNRAS}%
\def\prd{Phys.~Rev.~D}%
\def\nat{Nature}%
\begin{document}

\title{Evidences of an innermost stable bound orbit predicted by general relativity 
 from the amplitude of the twin-peak quasi-periodic oscillations}

\author{C. German\`a}
\email{claudio.germana@gmail.com}
\email{claudio.germana@ufma.br}
\author{R. Casana}
\affiliation{Departamento de F\'isica, Universidade Federal do Maranh\~ao, 
S\~ao Lu\'is, MA, Brazil}

\date{\today}

\begin{abstract}
The twin-peak high-frequency quasi-periodic oscillations (HF QPOs), observed
 in the power spectra of low-mass X-ray binaries, might carry relevant clues
 about the physics laws reigning close to a compact object. Their
 frequencies are typical of the orbital motion time-scales a few gravitational
 radii away from the compact object. 
 The aim of the manuscript is to propose an intuitive model explaining that the
 energy carried by the lower HF QPO can be related to differences of
 potential energy released by clumps of plasma spiraling in a curved space-time.
 Our model provides estimates on both the size of clumps of matter that can
 survive to the strong tidal force and energy loaded by tides on the clump. 
 We have also obtained some constraints on the mechanical properties of the
 plasma orbiting into the accretion disk. We note that the systematic behavior
 of the emitted energy as function of the central frequency of the lower HF QPO,
 observed in several sources with a neutron star, might give clues
 related to an innermost stable bound orbit predicted by the General
 Relativity theory in strong field regime.
\end{abstract}



\maketitle

\section{Introduction}\label{sec1}
Since their discovery \cite{1996ApJ...469L...1V}, the twin-peak high-frequency 
 quasi-periodic oscillations (HF QPOs) observed in the power
 spectra of low-mass X-ray binaries (LMXBs) have attracted much attention.
 The HF QPOs have frequencies that correspond to the time-scales of the
 orbital motion close to the compact object (milliseconds). They are believed to be produced by orbiting
 inhomogeneities of the density of the accretion disk, often referred as to clumps/blobs of matter 
 \cite{1998ApJ...508..791M,1999ApJ...524L..63S,2004ApJ...606.1098S,2014MNRAS.439.1933B}, 
 or to oscillation modes of the accretion disk \cite{2001ApJ...559L..25W,2009ApJ...704...68C,2012ApJ...752L..18W}.
Twin-peak HF QPOs might be potential probes to disclose the imprints produced by the
 orbiting matter in a strongly curved space-time, e.g. the signature of an innermost stable 
 circular orbit (ISCO) predicted by General Relativity (GR) in strong field 
 \cite{1990ApJ...358..538K,1998ApJ...500L.171Z}, as well as the modulations of the X-ray flux 
 by the precession of the orbits
 (see Refs.~\cite{2003ASPC..308..221L,2004astro.ph.10551V} for review). 
 HF QPOs can also be used
 to constrain fundamental quantities of the compact object, 
 such as its mass $m$ and angular momentum $a$ \cite{1972ApJ...178..347B}. 
 Attempts to estimates such quantities in systems with a neutron star 
 (NS LMXBs) were done soon after the discovery of HF QPOs 
 \cite{1999PhRvL..82...17S,1999ApJ...526..953K}. Most recently, 
\citet{2014MNRAS.437.2554M,2014MNRAS.439L..65M} gave measurements of both $m$ 
 and $a$ of black hole (BH) LMXBs. The observed
 frequencies of the HF QPOs were linked to the  relativistic frequencies as in
 the relativistic precession model (RPM; Ref.~\cite{1999ApJ...524L..63S}), where 
 the upper peak in frequency of the twin-peak HF QPOs is linked to the Keplerian 
 frequency $\nu_{k}$ of the orbiting matter, while the lower peak to 
 the periastron precession frequency $\nu_{p}$ of the orbit. Different
 interpretations \cite{1999ApJ...526..953K} and models linking $\nu_{k}$ 
 to the lower HF QPO \cite{1999ApJ...522L.113O,1999ApJ...518L..95T,2003ApJ...584L..83M} 
 were also proposed.

The unprecedented possibility of constraining the space-time around a compact 
 object through HF QPOs and the idea of preferred orbital radii where HF QPOs would be produced have 
 stimulated several works based on resonance mechanisms. 
 Resonance between relativistic epicyclic and Keplerian frequency of 
 the orbiting matter has been proposed to justify the observed 
 3:2 ratio of the twin-peak HF QPOs 
 \cite{2001A&A...374L..19A,2005AN....326..830R,2006A&A...451..377H,2013A&A...552A..10S}.    
 Other resonance mechanisms propose forced oscillations in the accretion disk 
 induced by the magnetic field of the 
 neutron star \cite{2005A&A...439..443P}, or resonance because of coupling between 
 the oscillation modes of the orbiting matter in the disk and the spin of the neutron star 
 \cite{2004ApJ...603L..89K,2004ApJ...603L..93L,2005A&A...439L..27P,2005A&A...443..777P,2009ApJ...694..387M}.
 Also, substantial work was done to test the models 
 with the observed data and constrain both $m$ and $a$ of the compact object  
 \cite{2011ApJ...726...74L,2012ApJ...760..138T,2014AcA....64...45S}.

Beside being potential probes to study the motion of matter in a strongly
 curved space-time, works were devoted to illustrate other properties of the HF QPOs. 
\citet{2013MNRAS.433.3453D,2013ApJ...770....9B} reported time-lags between photons from twin-peaks
 as collected in different energy bands in NS LMXBs. Such lags were previously also noted in 
 Ref.~\cite{1999ApJ...514L..31K}. Thanks to the larger sample of collected data,
 \citet{2013MNRAS.433.3453D,2013ApJ...770....9B} were able to get finer results. These information
 may shine light on the emission mechanisms in such extreme environment 
\cite{1999ApJ...519L..73F}.

\citet{2006MNRAS.370.1140B,2011ApJ...728....9B,2006MNRAS.371.1925M} analyzed
 the coherence 
 $Q$ of the twin-peaks, defined as $Q=\nu/\Delta\nu$, where $\nu$ is the
 central frequency of the peak and $\Delta\nu$ its width. In several NS LMXBs
 the $Q$-factor  of the lower HF QPO displays a steep increase and then an
 abrupt drop as function of the central frequency $\nu$ of the peak. It was
 proposed that the abrupt drop of the $Q$-factor might be caused by the
 approach of the oscillation to ISCO \cite{2006MNRAS.370.1140B}, thus a
 geometry-related effect. However, this interpretation was debated in
 Refs.~\cite{2006MNRAS.371.1925M,2007MNRAS.376.1139B}.
 If confirmed, a signature of ISCO is of relevant importance, because it is 
 predicted by GR in strong field regime. In Refs.~\cite{2006MNRAS.370.1140B,2011ApJ...728....9B,2001ApJ...561.1016M,2006MNRAS.371.1925M}
 is reported the amplitude of the twin-peak
 HF QPOs as function of their central frequency. Like the quality factor $Q$
 the behavior of the amplitude (the energy carried by the QPOs) of the lower 
 HF QPO displays a
 systematic trend with an increase and then a slightly steeper decrease for
 increasing central frequency of the peak. The amount of energy released is
 up to $\sim20$ \% the total flux of the source in NS LMXBs ($10^{34}-10^{36}$
 erg/s, from Atoll to Z-sources \cite{2006MNRAS.371.1925M}).
 But, how this huge quantity of energy is produced is not yet understood.

Matter orbiting close to a compact object experiences a strong tidal
 force. The effects of the encounter of stars with massive black holes were
 studied by a certain number of authors 
\cite{1985LAstr..99..429L,1988Natur.333..523R,1993ApJ...410L..83L,
2005ApJ...625..278G,2014ApJ...783...23G} 
 and are of interest for the flaring activity observed at the center of galaxies \cite{2012Natur.485..217G}. 
 \citet{2012Sci...337..949R} reported a QPO in the X-ray flux coming from the tidal disruption of a star 
 by a supermassive 
 black hole. Most recently, \citet{2014MNRAS.444...93D} linked to the tidal disruption of a planet by a white dwarf 
 the X-ray flaring activity observed in the source IGR J17361-4441. Subsequent analysis revealed a  
 QPO in the X-ray flux produced by such event \cite{2014A&A...570L...2B}.\\ 
The work done by tides is a significant source of energy
 already in our solar system. The volcanism in the Jupiter's moon Io is
 sustained by tides. Thus, it could be relevant to take into account the
 gravitational energy extracted through tides by the orbiting matter close to
 the compact object in LMXBs. \citet{2009A&A...496..307K}
 modeled the signal from the tidal disruption of small satellites orbiting a
 Schwarzschild black hole. \citet{2009AIPC.1126..367G} showed
 a simulated power spectrum with such code. For the first time is reported a
 power spectrum alike to the observed ones in LMXBs, both the power law and   
 twin-peaks are reproduced.

The shape of the amplitude of the lower HF QPO, as function of its 
 central frequency
 (fig.~1 in Ref.~\cite{2001ApJ...561.1016M}, fig.~3 in 
 Ref.~\cite{2006MNRAS.370.1140B} and
 fig.~2 in Ref.~\cite{2006MNRAS.371.1925M}), is typical of the potential energy
 released by
 matter spiraling in a curved space-time. We believe that it might be
 worthwhile to investigate such issue. In the following we concentrate on the
 lower HF QPO. We show for the first time that the energy carried by the lower
 HF QPO can be accounted for by the gravitational energy extracted by spiraling
 clumps of matter undergoing the work done by the strong tidal force. We make
 use of the Schwarzschild potential as first-order approach to the problem and check our 
 calculations also in the Kerr metric. We
 show that the Schwarzschild/Kerr gravitational energy extracted over different
 regions of the space-time reproduces the observed bell-shaped behavior of the
 amplitude of the lower HF QPO 
\cite{2006MNRAS.370.1140B,2011ApJ...728....9B,2006MNRAS.371.1925M,2001ApJ...561.1016M}. 
 The physical mechanism here proposed suggests that the 
 behavior may be an effect related to an innermost marginally stable orbit 
 predicted by GR in strong field regime, below which no stable orbital motion can exist
 \cite{1990ApJ...358..538K}. Because in theory such an orbit can have a generic 
 eccentricity $e$ \cite{1994PhRvD..50.3816C}, in a more general way we refer to it as to innermost
 stable bound orbit (ISBO) instead of ISCO.

The manuscript is organized as follow. In Section~\ref{sec2} we review the main features of
 the Schwarzschild potential and give new hints on a possible way to reveal a signature of ISBO. 
 Section~\ref{sec3}
 and Section~\ref{sec4} provide constrains on the maximum size allowed by tides of orbiting 
 clumps of solid-state and
 plasma-state matter, respectively. In Section~\ref{sec5} we estimate both the energy loaded by tides per 
 periastron passage on an
 orbiting clump of plasma and the amount of gravitational energy that would be emitted 
 after tidal disruption of the clump. 
 Section~\ref{sec6} shows our results in the Kerr metric. In Section~\ref{sec7} we comment on other  
 factors that might affect a signature of ISBO and show (Section~\ref{sec72}) an interpretation accounting 
 for the observed coherence of the lower HF QPO as function of the luminosity 
 of the source. Section~\ref{sec8} summarizes the conclusions.      

\section{Schwarzschild potential shape and ISBO}\label{sec2}

We are interested in estimating differences of potential energy released by
 matter spiraling between the orbits in a curved space-time. For 
 self-consistency we review some of the well known features of the
 gravitational potential in the Schwarzschild metric.
 Most importantly, we highlight new hints about a way of
 revealing a signature of ISBO.
 
The familiar radial effective Schwarzschild potential per unit mass $\mu$ of
 an orbiting particle reads \cite{1994PhRvD..50.3816C}  
\begin{equation}\label{eq1}
V_{eff}=1-\frac{2m}{r}-\frac{2m\tilde{L}^{2}}{r^{3}}+\frac{\tilde{L}^{2}}{r^{2}}
\end{equation}
where $m$ is the mass of the compact object and $\tilde{L}$ is the angular 
 momentum\footnote{The quantities here reported are expressed in geometric units 
 $G=c=1$, unless differently specified. The mass $m$ in geometric units is equal to 
 the gravitational radius of the compact object $r_{g}=GM/c^{2}$, where $M$ 
 is the mass of the compact object in SI units, $G$ the
 gravitational constant and $c$ the speed of light. For a $2\ M_{\odot}$ 
 neutron star $r_{g}\sim3$ km.} of the particle (per unit mass $\mu$) 
 on the orbit of semi-latus rectum $p$ and eccentricity $e$  
\begin{equation}
\tilde{L}=\left(\frac{p^{2}m^{2}}{p-3-e^{2}}\right)^{1/2}
\end{equation}
The quantity $p$ is defined such that the periastron of the orbit \mbox{$r_{p}=pm/(1+e)$} and for bound stable orbits \mbox{$p\ge6+2e$} 
 \cite{1994PhRvD..50.3816C}. The second
 and third term in (\ref{eq1}) refer to the gravitational potential in the
 Schwarzschild metric, while the forth term is the centrifugal potential. The
 attractive term $\propto1/r^{3}$ is relevant close to the compact object,
 i.e. when
 $r\sim r_{g}$ and causes the precession of elliptical orbits \cite{1916AbhKP1916..189S}. 

\begin{figure}
\includegraphics[width=0.47\textwidth]{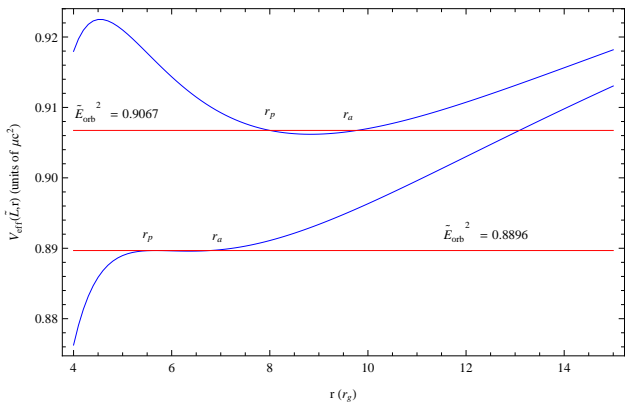}
\caption{Effective Schwarzschild potential as function of the radial
 coordinate $r$ for two different orbits of given specific angular momenta $\tilde{L}$ (blue curves). The periastra of the orbits are $r_{p}=5.6\ r_{g}$ (ISBO) and \mbox{$r_{p}=8\ r_{g}$}. The apastra are at $r_{a}=6.8\ r_{g}$ and $r_{a}=9.8\ r_{g}$, respectively.
Red lines show the orbital energy $\tilde{E}_{orb}^{2}$.}\label{fig1}
\end{figure}
Fig.~\ref{fig1} shows the known shape of the Schwarzschild potential for two
 different orbits. The blue lines draw the potential well for orbits of
 eccentricity $e=0.1$ and periastra $r_{p}=8\ r_{g}$ (top blue curve) and the
 ISBO $r_{p}=5.6\ r_{g}$ (bottom blue curve)\footnote{For an orbit with $e=0$ (ISCO) $r=6\ r_{g}$.}. The square of the orbital
 energy $\tilde{E}_{orb}^{2}$ (in units of mass $\mu$) is reported (red
 lines) and reads \cite{1994PhRvD..50.3816C}
\begin{equation}\label{eq3}
\tilde{E}_{orb}^{2}=\frac{\left(p-2-2e\right)\left(p-2+2e\right)}{p\left(p-3-e^{2}\right)}
\end{equation}
The abscissas of the crossing points between $\tilde{E}_{orb}^{2}$ and 
$V_{eff}$ are the periastron $r_{p}$ and the apastron $r_{a}$ of the orbit
 and come from the solution of the equation 
$\left(dr/d\tau\right)^{2}=\tilde{E}_{orb}^{2}-V_{eff}=0$ 
\citep{1994PhRvD..50.3816C}.    
The ISBO at $r_{p}=5.6\ r_{g}$ has a less deep minimum
 than the orbit at $r_{p}=8\ r_{g}$. It is because the term $\propto1/r^{3}$
 in (\ref{eq1}) starts dominating.
 
\begin{figure}
\includegraphics[width=0.47\textwidth]{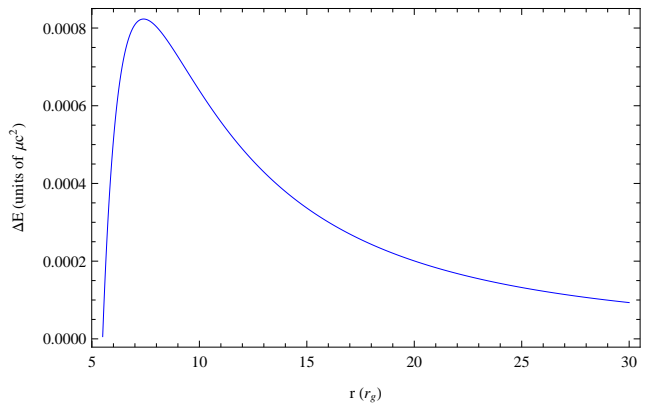}
\caption{Difference of potential energy at periastron between two orbits 
with periastra $0.1\ r_{g}$ away. The energy expected to be released draws 
a bell-like shape, decreasing towards ISBO ($r_{p}=5.6\ r_{g}$) and displaying
 its maximum in the range of radii $r_{p}\sim6.5-10\ r_{g}$.}\label{fig2}
\end{figure}

The flattening of the minimum of the potential is a well known feature of
 GR \cite{misner}. Besides the precession of the orbits that has already been
 measured in the
 case of Mercury (but in a weak field regime), the observational implications
 of the term $\propto1/r^{3}$ in (\ref{eq1}) in a strong field regime could
 be unmasked by the potential energy released by spiraling matter close the 
 compact object. It is well known from the accretion theory 
\cite{2002apa..book.....F} that if matter accretes from an orbit at $r=a\ r_{g}$
 and ends up onto an inner orbits at $r=b\ r_{g}$ 
 (b $<$ a), the energy released is the difference of potential energy
 between the two orbits. Such spiraling motion may be caused because of,
 e.g., the removal of orbital energy by tides acting on the clump of matter \cite{2008A&A...487..527C}.\\ Fig.~\ref{fig2} 
shows a possible and simple way that could disclose a signature of ISBO. It
 depicts the behavior of the difference of the Schwarzschild
 potential, taken at periastron, between two orbits with periastra, e.g., 
 $0.1\ r_{g}$ away. The orbits have a low eccentricity $e=0.1$. The range of 
 periastra is from $r_{p}=30\ r_{g}$ to the ISBO at $r_{p}=5.6\ r_{g}$. 
 Such difference of potential energy between the two
 orbits, expected to be released, draws a bell-like shape: it increases up to $\sim7.5\ r_{g}$ and then
 decreases up to ISBO. This characteristic shape is a consequence of the term
 $\propto1/r^{3}$ in (\ref{eq1}). In a Newtonian potential the differences of
 potential keep increasing in absolute value towards the inner regions. Thus,
 an observational evidence of the term $\propto1/r^{3}$ in (\ref{eq1}) and,
 therefore of ISBO, might be through these energetic considerations. We stress
 that these arguments have not yet been highlighted in the literature. 

In Refs.~\cite{2001ApJ...561.1016M,2006MNRAS.371.1925M,2006MNRAS.370.1140B} it is shown 
 that the energy (the amplitude) carried by
 the lower HF QPO in NS LMXBs, as function of its central frequency,
 draws a bell-shaped behavior like that around the maximum of fig.~\ref{fig2}. 
 The central frequency of the lower HF QPO is typical of the orbital
 motion within $r\sim6.5-10\ r_{g}$. It may be worthwhile to investigate
 whether the observed behavior originates from differences of potential energy
 released by spiraling matter. 

\section{Size allowed by the strong tides of orbiting clumps of matter: 
the solid-state case}\label{sec3}

The arguments in Section~\ref{sec2} give general clues about the overall
 pattern of the potential energy (in units of rest-mass energy) released by
 spiraling matter between orbits $0.1\ r_{g}$ away over
 different regions of the space-time, to compare it to the observed pattern
 reported
 in fig.~1 in Ref.~\cite{2001ApJ...561.1016M}, fig.~3 in Ref.~\cite{2006MNRAS.370.1140B} and fig.~2 in Ref.~\cite{2006MNRAS.371.1925M}. We
 investigate whether this mechanism could actually account for the observed
 energy carried by the lower
 HF QPO ($10^{34}-10^{36}$ erg/s, from Atoll to Z-sources 
 \cite{2006MNRAS.371.1925M}). We need to give an estimation on the mass of the
 clumps of matter. The strong tidal force around a compact object sets an upper
 limit on their size.

Consider a spherical clump of matter of radius $R$ and its cup of height
 $h=R/10$ at the  distance $\sim R$ from the center of the clump. The volume
 of the cup is $V= \pi h^{2}(R-h/3)$, thus its mass is $\mu'=\rho V$, where 
 $\rho$ is the density of the material. The effective gravitational force
 acting on the cup at distance $r-R$ from the compact object is 
\begin{eqnarray}\label{eq4}
F_{eff}&=&\mu'c^{2}\left(\frac{dV_{eff}}{dr}\right)_{(r-R)}\\ \nonumber
&=&\mu'c^{2}\left(\frac{2m}{(r-R)^{2}}+\frac{6m\tilde{L}^{2}}{(r-R)^{4}}-\frac{2\tilde{L}^{2}}{(r-R)^{3}}\right) 
\end{eqnarray}     
where $V_{eff}$ is from (\ref{eq1}) and $c^{2}$ has been inserted to convert
 from geometric to SI units. The same cup placed on the opposite side of the
 clump at $r+R$ would undergo the force 
$F_{eff}=\mu'c^{2}\left(\frac{dV_{eff}}{dr}\right)_{r+R}$, therefore the
 tidal force over the clump of matter is (see also Ref.~\cite{1996Icar..121..225A})
\begin{eqnarray}\label{eq5}
F_{T}&=&\mu'c^{2}\left[\left(\frac{dV_{eff}}{dr}\right)_{(r-R)}-\left(\frac{dV_{eff}}{dr}\right)_{(r+R)}\right]\\ \nonumber
&\approx&\mu'c^{2}2R\left(\frac{d^{2}V_{eff}}{dr^{2}}\right)_{r}
\end{eqnarray}
where $\approx$ indicates that we used a Taylor sum in $R/r$ up to the first
 order since we expect $R<<r$. The tidal force can not be larger than the
 ultimate tensile strength $\sigma$ of the material\footnote{In our solar
  system there are moons orbiting the parent planet below their gravitational
 Roche radius. They are held together by  electrochemical bonds stronger than
 their self-gravity. Metis and Pan are moons orbiting below their gravitational
 Roche radius.} (times the area $2\pi R h$
 of the cup), i.e. $F_{T}\leq2\pi R h\sigma$. From this inequality we can get
 some order of magnitude on the maximum radius $R$ allowed by tides
\begin{eqnarray}\label{eq6}
R&=&\left(10\left(1-\frac{1}{30}\right)^{-1}\frac{c_{s}^{2}}{c^{2}}\frac{\sigma}{Y}\times\right.\\ \nonumber
&&\left.\left(-\frac{2m}{r^{3}}+\frac{3\tilde{L}^{2}}{r^{4}}-\frac{12m\tilde{L}^{2}}{r^{5}}\right)^{-1}\right)^{1/2}
\end{eqnarray}
where we write the density $\rho=Y/c_{s}^{2}$, $Y$ is the Young's modulus of
 the material and $c_{s}$ the speed of sound in it. In the case of a 
 solid-state material, e.g. asteroids\footnote{Even if an asteroid would
 probably be vaporized by the strong X-ray field ($T\sim10^{7}$ K) far away
 the compact object, we consider this case for the sake of completeness.},
 iron/steel-like material has a typical ratio in the laboratory $\sigma/Y\sim5\times10^{-3}$ ($\sigma\sim10^{2}$ MPa, $Y\sim10^{2}$ GPa \citep{kalpa})
 and $c_{s}=5\times10^{5}$ cm/s. From (\ref{eq6}) the upper limit on the 
radius $R$ of a solid-state clump that can survive to tides, at a distance $r\sim6-10\ r_{g}$ from a $2\ M_{\odot}$ neutron star, is $\sim30$ cm 
(see also Ref.~\cite{2009NCimB.124..155K}). The rest-mass energy of a such
 sphere made of iron or steel \mbox{($\rho\sim8$ g/cm$^{3}$)} is 
 $\mu c^{2}\sim10^{27}$ erg, thus from fig.~\ref{fig2} only $\sim6\times10^{23}$
 erg can be emitted at the maximum of the bell-shape curve. This is too low for
 the observed energy of the lower HF QPO whose amplitude is some
 percent of the luminosity of the source, i.e. $10^{34}-10^{36}$ erg/s. Hence,
 solid-state clumps made of  iron/steel-like material (e.g. asteroids) are
 disfavored as sources of the lower HF QPO. 

\section{Size allowed by the strong tides of orbiting clumps of matter: 
the plasma case}\label{sec4}

The plasma material in the accretion disk has different mechanical properties
 than solid-state matter. However, it may raise some doubts thinking that
 clumps of plasma in the accretion disk can survive to the strong shear forces
 due to differential rotation, thus not able to produce the lower HF
 QPO. On the other hand, to our knowledge, detailed simulations showing that
 clumps
 of plasma can not produce QPOs in the power spectra are not reported in the
 literature. Instead, there is a numerical code 
\cite{2009A&A...496..307K} showing that orbiting clumps around a compact object can produce power spectra with twin-peak QPOs
 \cite{2009AIPC.1126..367G}. In this numerical code the clump is
 a collection of free particles. Each particle has its own geodesic. By means
 of ray-tracing techniques in
 the Schwarzschild metric the code shows that the signal from the clump,
 disrupted by both shear forces and tides, can produce twin QPOs. Most
 recently, \citet{2013MNRAS.430L...1G} argued that the timing law of the
 azimuth phase $\phi(t)$ on elliptical relativistic orbits could be the root
 for the
 appearance of multiple peaks in the power spectrum. 
\citet{2014MNRAS.439.1933B} modeled in detail the signal produced  
 by orbiting hot-spots. Although the issue of a lower HF QPO produced 
 by orbiting clumps in the disk may be of open debate, it is a viable
 interpretation. We should bear in mind that the observed central frequency
 of the lower HF QPO is typical of the orbiting matter close to the
 compact object \cite{1998ApJ...508..791M,1999ApJ...524L..63S,2001A&A...374L..19A,
2004ApJ...606.1098S,2014MNRAS.439.1933B}.
 Moreover, the numerical code in Ref.~\cite{2009A&A...496..307K} shows that 
\emph{already} clumps of free particles can produce a lower HF QPO \cite{2009AIPC.1126..367G}.  
 In reality, since the clumps of plasma in the disk might have some internal
 force, it is even
 more probable that they produce the lower HF QPO. Thus, it might be
 worthwhile to
 investigate the mechanical properties of clumps of plasma in the accretion
 disk.

We estimate the density of the inner part of the accretion disk by using
 the equation of the density of the disk reported in  
 Ref.~\cite{2002apa..book.....F}. For typical accretion rates in NS LMXBs
 of $\sim10^{18}$ g/s (equation 1.7 in 
 Ref.~\cite{2002apa..book.....F}; a Z source with luminosity $L=L_{Edd}\sim2.5\times10^{38}$
 erg/s \cite{2006MNRAS.371.1925M}, where $L_{Edd}$ is the Eddington limit), for a $2\ M_{\odot}$ neutron star and at a
 distance of $r\sim6-10\ r_{g}$, the density of the material in the disk is 
$\sim10$ g/cm$^{3}$, like that of iron.  
Because it is expected to be a highly ionized plasma, the speed of sound $c_{s}$ is not like that in iron materials. We write 
\cite{2002apa..book.....F}
\begin{equation}
c_{s}=\left(\frac{\gamma Z k T}{m_{i}}\right)^{1/2}    
\end{equation}
where $\gamma\sim5/3$ is the adiabatic index, $Z$ the charge state ($Z=1$ for
 a hot plasma), $m_{i}$ the ion hydrogen mass, $k$ the Boltzmann's constant.
 For typical temperatures in the inner part of the accretion disk of 
$T\sim4\times10^{7}$ K it turns $c_{s}\sim8\times10^{7}$ cm/s.\\ 
From (\ref{eq6}) the upper limit by tides on the size of the clump depends also on
 the ratio $\sigma/Y$. This quantity describes the
 malleability of the body to external deformations, or, following the Hooke's
 law ($\sigma=Y\epsilon$), it tells us how much deformable the material
 is before breaking. In the following we explore the idea of characterizing clumps of plasma
 through the ratio $\sigma/Y$. This way of proceeding might be justified by the results reported
 by \citet{2013Sci...339.1048C} who discovered 
 large structures in the
 accretion disk of an X-ray binary system.
 Such structures propagate through the disk. Thus, it is hard to think to a completely smoothed accretion disk, but rather it might be characterized by inhomogeneities propagating throughout it. 
If inhomogeneities of plasma exist, as reported in Ref. 
\cite{2013Sci...339.1048C}, they have to be held together by some internal force (e.g. Ref.~\cite{1989Ap&SS.158..205H}) against tides and differential rotation.
 Whatever the nature of this force is, its value per unit area displays an
 ultimate tensile strength of the material. Equivalently, such force can be
 characterized through the mechanical binding energy, defined as the energy
 required to disaggregate/separate a body \cite{chandra}. The binding energy
 per unit volume is the ultimate tensile strength $\sigma$
 \cite{2002aste.conf..463A}. 
 
In this context, it might be worth noting that some features of the accretion disk theory is
 still of open debate. \citet{2007MNRAS.376.1740K,2013MNRAS.431.2655K} pointed out the yet
 unclear understanding about the viscosity in accretion disk models. They investigated
 the disagreements between the viscosity parameter $\alpha$ required from observations and the
 one from simulations. \citet{2007MNRAS.376.1740K} remark that $\alpha$ is not
 unambiguously the viscosity in the Navier-Stokes equation but rather we can
 think it as the ratio of the stress to the rate of strain of the plasma. Its
 real physical meaning remains unclear.

\subsection{Constrain the ratio $\sigma/Y$ of clumps of 
plasma in the accretion disk}\label{sec42}
 
We estimate the ratio $\sigma/Y$ recalling the binding
 energy of a body. The gravitational binding energy of an uniform sphere of
 mass $M$ and radius $R$ is defined as the energy required to separate it in
 small parts and bring them at an infinite distance. The formula reads 
$E_{b}=3GM^{2}/5R$ \cite{chandra}. This is valid for a body of loose
 material held together by only gravity. We are dealing with
 clumps of hot-dense plasma, held together by some internal force. If $\sigma$
 is the ultimate tensile strength to be applied to break a specimen of such
 material, i.e. an energy per unit volume \cite{2002aste.conf..463A},
 the energy required to disaggregate a body of volume $V$ is the mechanical
 binding energy 
\begin{equation}\label{eq8}
E_{b}=\sigma V=\frac{4}{3}\pi R^{3}\sigma
\end{equation}  
Substituting $R$ from (\ref{eq6}) we obtain
\begin{eqnarray}\label{eq9}
\frac{\sigma}{Y}&=&\left(\frac{3}{4\pi}\left(\frac{1}{10}\right)^{3/2}\left(1-\frac{1}{30}\right)^{3/2}\frac{E_{b}}{Y}\left(\frac{c}{c_{s}}\right)^{3}\right.\times\\ \nonumber
&&\left.\left(-\frac{2m}{r^{3}}+\frac{3\tilde{L}^{2}}{r^{4}}-\frac{12m\tilde{L}^{2}}{r^{5}}\right)^{3/2}\right)^{2/5}
\end{eqnarray}
Accreting clumps of plasma once disrupted by tides should release an energy
 that is at least of the same order of $E_{b}$. Thus, such
 binding energy may be of the order of that emitted by the lower HF QPO. For a Z source, the amplitude of the lower HF QPO is $\sim2-4\%$ the luminosity of the source \cite{2006MNRAS.371.1925M}. We shall
 write $E_{b}\sim5\times10^{36}$ erg. Knowing the Young's modulus $Y=\rho c_{s}^{2}$
 for the clump of plasma, we get $\sigma/Y\sim70$. A ratio $\sigma/Y>1$ means a deformation at break as big as more than once the initial size of the body. This is the case for, e.g., an elastic rope. Thus, a $\sigma/Y\sim70$ means
 that the clump of plasma is far more deformable than solid-state iron
 materials. Because of this higher malleability to external  deformations,
 (\ref{eq6}) gives a bigger upper limit on the size of the structure of
 plasma that sustains tides, $R\sim6500$ m.

In Sec.~\ref{sec4} we mentioned that already clumps of free particles
 orbiting a compact object can  produce the lower HF QPO 
\cite{2009AIPC.1126..367G}. Here the clump of plasma is 
 held together by an internal force characterized by $\sigma$. We calculate
 whether the large upper limit of $R\sim6500$ m set by tides can sustain
 the strong shear forces due to differential rotation. We divide the
 structure into, say, 10 slices and calculate the differential velocity as 
\begin{equation}
dv=2\pi r\left(\frac{GM}{4\pi^{2}r^{3}}\right)^{1/2}-2\pi r'\left(\frac{GM}{4\pi^{2}r'^{3}}\right)^{1/2}
\end{equation}  
where $r$ is the radius of the orbit at the center of the clump, $r'=r+dh$ is
 the orbital radius of a contiguous slice. The shear force between two
 contiguous slices is  
\begin{equation}
F=\rho \nu_{k} dV dv
\end{equation} 
where $\rho\sim10$ g/cm$^{3}$ is the density of the plasma, $dV=2\pi R^{2} dh$
 is the volume of the slice and $\nu_{k}$ the Keplerian frequency. It turns
 out that such load is $F=10^{24}$ N. The ultimate tensile strength 
$\sigma\sim70 Y$ gives an opposite force keeping the slices together of $2\pi R^{2}\sigma\sim10^{25}$ N.

\section{Energy loaded by tides on orbiting clumps of plasma}\label{sec5}

With the limits on the size of the clump of plasma estimated above, we may
 now give constrains on both the energy that tides might deposit on the orbiting
 clump and the amount of removed orbital energy ($\ref{eq3}$) per periastron
 passage. This will give us information on the interval of orbits the clump
 spirals over before disrupting and constrain how much potential energy is
 released, to compare it to the observations. 

The rate of energy deposited by
 tides on the orbiting clump of plasma is \cite{1977ApJ...213..183P}
\begin{equation}
\frac{dE}{dt}=\int{dx^{3} \rho\ \mathbf{v} \cdot \bm{\nabla} V_{eff}}
\end{equation}
where \textbf{v} is the velocity of an element of the clump because of the
 perturbation by tides. We shall write $\left|\mathbf{v}\right|=dR/dt$, with $R$ radius of the
 clump. We turn the integral in time into an integral over the orbit,
 multiplying by $dt/dr$
\begin{equation}\label{eq13}
\frac{dE}{dr}=\int{dx^{3} \rho\ \frac{dR}{dr} \nabla V_{eff}}
\end{equation}  
The term $dR/dr$ simulates the deformation of the clump because of tides as
 function of the distance $r$ from the central object. On a slightly eccentric
 orbit $dR/dr$ oscillates because of the changing tidal force between periastron
 and apastron. At $r\sim8\ r_{g}$ the tidal force (\ref{eq5}) per unit
 area
 over a clump of plasma $R=3200$ m is $\sigma_{T}\sim10^{17}$ Pa, while the
 ultimate tensile strength of the material $\sigma=70Y\sim5\times10^{17}$ Pa.
 Thus, the clump may undergo some elastic phase before 
 breaking\footnote{Structures equal/larger than the limits constrained 
 in Sec~\ref{sec42} can not survive at all because of tides. On the other
 hand, structures smaller than $R\sim3000$ m would not be able to emit the
 required energy seen in the lower HF QPO in Z sources, i.e. 
 $\sim5\times10^{36}$ erg/s.}. The tidal stress $\sigma_{T}$ slightly
 oscillates (by a factor $\sim2$) from periastron to apastron on a time-scale
 \footnote{$\nu_{r}$ is the relativistic radial frequency of the orbit. In the
 Schwarzschild metric around a $2\ M_{\odot}$ neutron star $\nu_{r}\sim350$ Hz
 at its maximum  ($r\sim8\ r_{g}$).} $t=1/\nu_{r}\sim3\times10^{-3}$ s.\\ 
Because $\sigma_{T}<\sigma$ to get order of magnitudes we treat the clump as
 simple elastic body deformed by tides. We make use of the Hooke's law $d\sigma=Y\ dR/R$,
 i.e. an infinitesimal variation of the load causes an infinitesimal relative
 deformation proportional to the Young's modulus $Y$ of the material. The
 term $dR/dr$ in (\ref{eq13}) turns into $(R/Y) d\sigma/dr$ and since the
 load $\sigma$ is the tidal force (\ref{eq5}) per unit area
\begin{equation}\label{eq14}
\frac{dE}{dr}=\frac{1}{4\pi R\ Y}\int{dx^{3} \rho\ \frac{dF_{T}}{dr}\nabla V_{eff}}
\end{equation}
\begin{figure}[!top!]
\includegraphics[width=0.47\textwidth]{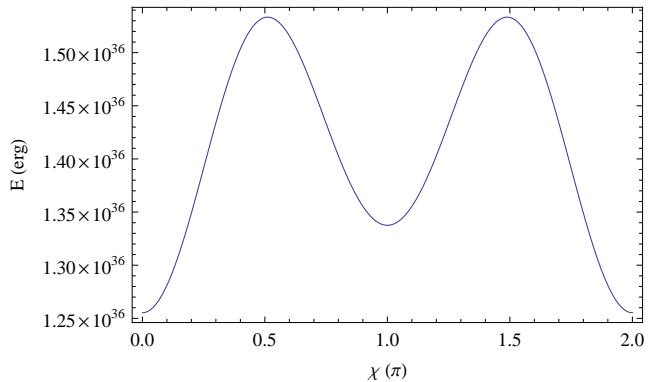}
\caption{Energy loaded by tides on a clump of plasma of radius $R=3200$ m as function of the
 radial phase $\chi$ of the orbit.}\label{fig3}
\end{figure}
The tidal force is constant over the clump at a given $r$. The integral over
 the volume, after transforming it in spherical coordinates, reads 
\mbox{$\approx4\pi R^{3}(dV_{eff}/dr)_{r}$} in the approximation $R<<r$.
 Substituting (\ref{eq5}) into (\ref{eq14}) and integrating over the radius
 of the orbit, we obtain an estimation on the energy loaded by tides on the
 clump
\begin{eqnarray}\label{eq15}
E\left(r\right)&=&\frac{3 \pi \rho c^{4} R^{6}}{6.6\times10^{3} c_{s}^{2} r^{9}}\times \\ \nonumber
&&\left(20\tilde{L}^{4}\left(21mr-35m^{2}-3r^{2}\right)-42m^{2}r^{4}+\right.\\ \nonumber
&&\left.-15 \tilde{L}^{2}mr^{2}\left(26m-7r\right)\right)
\end{eqnarray}
We insert into (\ref{eq15}) the parametrized radius \citep{1994PhRvD..50.3816C}
\begin{equation}
r\left(\chi\right)=\frac{p m}{1+e\cos\left(\chi\right)}
\end{equation} 

\begin{figure}[top!]
\includegraphics[width=0.47\textwidth]{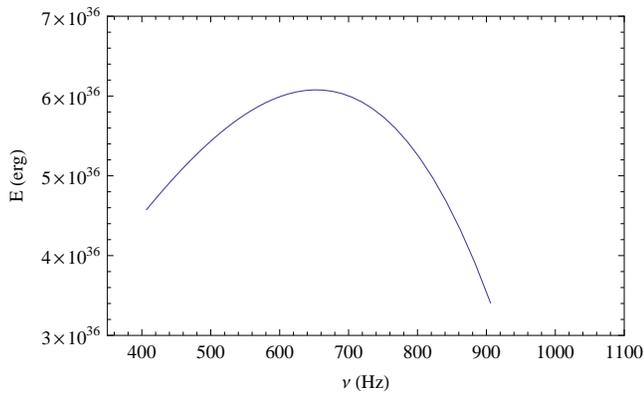}
\caption{Difference of potential energy at periastron between two orbits with periastra $\sim1700$ m away as function of the orbital frequency. 
To compare it to the data see fig.~1 in Ref.~\cite{2001ApJ...561.1016M}, fig.~3 in Ref.~\cite{2006MNRAS.370.1140B} and fig.~2 in 
Ref.~\cite{2006MNRAS.371.1925M}.}\label{fig4}
\end{figure}

Fig.~\ref{fig3} shows the behavior of the energy loaded by tides\footnote{Note
 that the order of magnitude estimated here \mbox{$E(r)\sim0.12\%\ \mu c^{2}$}
 agrees with that from the formalism described in  
 Ref.~\cite{2005ApJ...625..278G} for a star disrupted by a supermassive black
 hole.} (\ref{eq15})
 on the clump of plasma as function of the radial phase $\chi$. The curve is
 symmetric with respect to the apastron at $\chi=\pi$. The orbit has the
 periastron at $8\ r_{g}$ and eccentricity $e=0.1$, around a $2\ M_{\odot}$ neutron 
 star\footnote{All over the window of periastra $\sim5.6-10\ r_{g}$ the energy
 deposited on the clump $R=3200$ m is $\sim10^{36}$ erg.}.\\ 
Such energy is transferred from orbit to internal energy through tides 
\cite{1977ApJ...213..183P, 2008A&A...487..527C}. A removal of orbital
 energy of $\sim10^{36}$ erg from (\ref{eq3}) gives a spiral motion of 
 $0.4\ r_{g}\sim1200$ m after the first periastron passage on these orbits with
 low $e=0.1$. The tidal wave propagates through the clump on a time-scale 
 $\tau=2R/c_{s}$. The radius of the fraction of the clump disrupted between two 
 periastron passages is $R_{l}=t c_{s}/2\sim1100$ m. On the second periastron passage
 the radius of the clump might be smaller ($R\sim2000$ m), the energy deposited by tides 
 (\ref{eq15}) now is $E(r)\sim10^{35}$ erg causing a spiral motion of $0.1\ r_{g}\sim300$ m.
 On the third periastron passage what is left has a radius $R\sim1000$ m $< R_{l}$. 
 The energy deposited
 by tides on such relic at each periastron passage is 
 $E(r)\sim10^{33}$ erg causing a spiral motion of $0.01\ r_{g}\sim30$ m. The binding energy
 (\ref{eq8}) is $E_{b}\sim10^{34}$ erg. Thus, the number of periastron passages before
 disrupting may be $\sim10$.
 In total the clump might make $\sim12$ periastron passages. The number of 
 Keplerian turns is\footnote{Around the region 
 $r\sim8\ r_{g}$ the ratio of the 
 relativistic Keplerian frequency $\nu_{k}$ to the radial one $\nu_{r}$ is $\nu_{k}$/$\nu_{r}\sim2$. Thus, if the number of periastron passages is $n$, the
 number of Keplerian turns is $2n$.} $\sim24$. The difference of potential energy
 between the orbits is emitted by such an event on a time-scale 
 $\tau'=$12/(350 Hz)$\sim0.034$ s. The coherence as defined in 
 Section~\ref{sec1} $Q=\nu/\Delta\nu$ in this case is $Q=\nu_{k}\tau'\sim24$,
 corresponding to the number of Keplerian turns.\\ 
In total, the initial clump with $R=3200$ m has spiraled over an interval of
 orbits $\sim0.6\ r_{g}\sim1700$ m. Fig.~\ref{fig4} shows the differences of
 potential energy, at periastron,
 released  between two  orbits with periastra \mbox{$\sim1700$ m} away as
 function of the orbital frequency\footnote{We calculated the orbital frequency for an orbit
 with $e=0.1$ by means of the integrals reported in Ref.~\cite{1994PhRvD..50.3816C} describing both the
 azimuth phase $\phi(\chi)$ and coordinate time $t(\chi)$ in the Schwarzschild metric, for
 generic orbits of eccentricity $e$. We resolved them by applying a Taylor sum around $e=0$ 
 \cite{2013MNRAS.430L...1G}.} 
 around a $2\ M_{\odot}$
 Schwarzschild compact
 object (the curve spans over \mbox{$r\sim10-6\ r_{g}$}). It displays the
 characteristic bell-shape centered at $\sim700-800$ Hz\footnote{For a $1.8\ M_{\odot}$ compact
 object the curve in fig.~\ref{fig4} is centered around $800$ Hz.} seen in the observations 
\cite{2001ApJ...561.1016M,2006MNRAS.370.1140B,2006MNRAS.371.1925M,2011ApJ...728....9B}.
 The emitted energy matches that carried by the lower HF QPO in Z
 sources \cite{2006MNRAS.371.1925M} and corresponds to $\sim0.5\%\ \mu c^{2}$.
 The figure corresponds to the maximum of
 the overall bell-shape in fig.~\ref{fig2}. Because the bell-shape in 
 fig.~\ref{fig4} is typical of the observed amplitude behavior of the
 lower HF QPO rather than the upper
 one, this study may add one more clue: it suggests that  the
 lower peak of the observed twin-peaks corresponds to the Keplerian
 frequency. 
This is in agreement with the results from the numerical
 code on tidal disruption \cite{2009AIPC.1126..367G}, 
 where the lower HF QPO corresponds to the Keplerian modulation $\nu_{k}$
 while the upper HF QPO to the modulation at $\nu_{k}+\nu_{r}$. Such
 conclusion is also reported in Ref.~\cite{2013MNRAS.430L...1G}.
 Models proposing other mechanisms to produce HF QPOs and assuming the lower 
 HF QPO to be $\nu_{k}$ have been previously proposed 
 \cite{1999ApJ...522L.113O,1999ApJ...518L..95T,2003ApJ...584L..83M}, 
 differently than in  
 Refs.~\cite{1998ApJ...508..791M,1999ApJ...524L..63S,2014MNRAS.437.2554M,2014MNRAS.439L..65M}
 in which the upper HF QPO is linked to $\nu_{k}$. 

 The lower HF QPO is observed to range $\sim600-950$ Hz \cite{2001ApJ...561.1016M,2006MNRAS.370.1140B,2006MNRAS.371.1925M,2011ApJ...728....9B}.
 From fig.~\ref{fig4}  we deduce that the clustering might be because
 the difference of potential energy released by such random events between near orbits has its
 maximum in the region of the space-time $r\sim10-6\ r_{g}$ 
(see also fig.~\ref{fig2}).    
The steeper decrease in fig.~\ref{fig4} is drawn by the approach to ISBO 
(see also fig.~\ref{fig2}). Hence, the observed behavior of the amplitude of
 the lower HF QPO as function of its central frequency may give clues related
 to the ISBO predicted by GR in a strong field
 regime ($r\sim r_{g}$). 

\section{The Kerr metric case}\label{sec6}
We performed our calculations in the case of a rotating compact object, whose 
 space-time around it is described by the Kerr metric \cite{1963PhRvL..11..237K}. The effective  
 radial gravitational potential $V_{eff, Kerr}$ is such \cite{1972ApJ...178..347B} that for a 
 non-rotating compact object ($a=0$, $a$ is the angular momentum per unit mass of the compact
 object) we get (\ref{eq1}) 
\begin{eqnarray}\label{eq17}
V_{eff,Kerr}&=&1-\frac{2m}{r}+\frac{a^{2}}{r^{2}}+\left(1+\frac{2 m}{r}\right)\frac{a^{2}\tilde{E}_{orb}^{2}}{r^{2}}+ \\ \nonumber
&+&\frac{4 am\tilde{E}_{orb}\tilde{L}}{r^{3}}+\left(1-\frac{2m}{r}\right)\frac{\tilde{L}^{2}}{r^{2}}
\end{eqnarray}
with $0\leq a\leq m$ and $\tilde{E}_{orb}$, $\tilde{L}$ orbital energy and 
 specific angular momentum of the orbiting particle on a generic circular  
 orbit in the Kerr metric \cite{1972ApJ...178..347B}.
\begin{figure}
\includegraphics[width=0.47\textwidth]{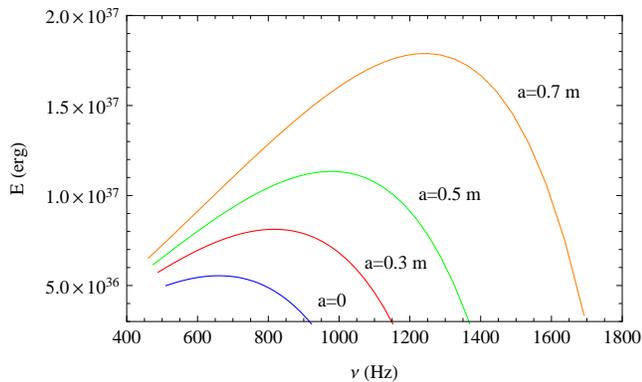}
\caption{Differences of Kerr/Schwarzschild potential energy expected to be released by an 
 orbiting clump of plasma of radius $R=3200$ m after being disrupted by tides. The curves refer to 
 different values of the Kerr parameter $a$.}
\label{fig5}
\end{figure}   

We note that the Kerr parameter $a$ lowers the maximum radius $R$ of the clump of 
 plasma allowed by tides: 
 for an orbit at $8\ r_{g}$ $R\sim7000$ m\ \ \footnote{This value agrees to that 
 from (\ref{eq6}) for an orbit with $e=0$.} for $a=0$, $R\sim6500$ m for $a=0.9m$. 
 The work done by tides per periastron passage remains of the same order of that 
 obtained in the Schwarzschild metric (equation (\ref{eq15}), $E(r)\sim10^{36}$ erg). Thus, the clump
 with 
 size $R=3200$ m as in Sec.~\ref{sec5} makes an almost equal number of turns before 
 disrupting, spiraling over an interval of orbits of $\sim0.6\ r_{g}\sim1700$ m. Fig.~\ref{fig5} 
 shows the difference of Kerr potential energy released for different values of $a$ between orbits 
 $\sim1700$ m away. We see that such mechanism is more efficient for increasing $a$: the amplitude of the bell-shape curve 
 increases. The effect of $a$ is also 
 to move the ISBO to the inner regions, so that the orbital frequency extends 
 to higher values. The
 figure refers to a $2\ M_{\odot}$ compact object. For a $2.5\ M_{\odot}$ each curve in the figure is almost shifted 
 to lower frequencies by $\sim200$ Hz. It is known that different couples 
 ($m,a$) could describe the  same frequency at
 a given orbital radius \cite{2012ApJ...760..138T}. 
 Instead the value of the mass has no effects on the amplitude of the bell-shape curves in 
 fig.~\ref{fig5}.

\section{Other factors affecting the amplitude of the lower HF QPO}\label{sec7}

We advise caution about the behavior shown in figs.~\ref{fig4}, \ref{fig5} and seen in the
 observations \cite{2001ApJ...561.1016M,2006MNRAS.370.1140B,2006MNRAS.371.1925M,2011ApJ...728....9B}
 as candidate to disclose a signature of ISBO. We discuss below other factors
 that may affect the amplitude of the lower HF QPO. 

\subsection{The dependence of the amplitude on the energy}

The amplitude of the lower HF QPO measured at different energy channels keeps
 increasing towards hard X-ray. At $\sim20-30$ keV the amplitude of the lower
 HF QPOs is $\sim20\%$,
 while at $\sim2$ keV is less than $5\%$ \cite{1996ApJ...469L..13B,2001ApJ...561.1016M}. 
The lower HF QPO is usually stronger in the hard state of the source, where the
 energy spectrum is dominated by a power law. This means that  
 the thermal emission from a Keplerian disk can not justify its nature. 
 A corona of hot electrons scattering to
 higher energy the seed photons from the disk has been proposed 
 \cite{2000ApJ...544L.119D, 2001ApJ...549L.229L,2001ApJ...554...49D,2013MNRAS.432.1144S}.  
On the other hand, \citet{2005tsra.conf..511S} 
noted that this mechanism may not amplify the modulation, but rather would
 lower the amplitude of the modulation at high energies, since the scattering
 of photons by hot-electron would smooth the oscillation from the disk.\\
 Alternatively to the corona scenario, the modeling in 
 Ref.~\cite{1995ApJ...455..623C}
 shows that sub-Keplerian motion in the inner part of the accretion  
 disk ($r\lesssim20\ r_{g}$) might cause a postshock region with hard X-ray 
 emission. Hard X-ray 
 emission also might come from the converging bulk motion of the accreting matter. 
 In Ref.~\cite{2003MNRAS.342..274M} 
 it is shown that in such sub-Keplerian region the specific angular 
 momentum of the accretion flow 
 can also behave in a Keplerian way. Therefore, Keplerian imprints 
 like that described in our manuscript may be seen in the observed hard state.\\      
\citet{2011RAA....11..631Y} pointed out the idea that the hard X-ray spectrum 
 of the lower HF QPO
 might originate by synchrotron mechanisms from magnetized hot-spot. 
 Calculations on the radiation that would be emitted by a tidally 
 disrupted magnetized clump of matter were attempted in 
 Refs.~\cite{2009A&A...496..307K,2010AIPC.1205...30C} 
 and full magnetohydrodynamic simulations are demanded. Thus, the exact 
 energy emission mechanism of the lower HF QPO may
 remain of open debate and is far beyond the goal of this paper. This
 manuscript attempts to propose a mechanism justifying how the quantity of
 energy carried by the lower HF QPO would originate and not the energy spectrum
 of the emission, which depends on the exact nature of the energy emission
 mechanism. We are comparing our calculations to bolometric luminosity.
 Gravitational energy is extracted through tides and transformed into radiation
 through a given emission mechanism. 
 The bolometric luminosity emitted should be equal to the amount of
 gravitational energy extracted. Thus, the behavior in figs.~\ref{fig4}, \ref{fig5} and seen
 in the observations \cite
{2001ApJ...561.1016M,2006MNRAS.370.1140B,2006MNRAS.371.1925M,2011ApJ...728....9B}
 may still remain a candidate to disclose a signature of ISBO.

\subsection{The dependence of the amplitude on the luminosity of the source}\label{sec72}

It is known that the amplitude of the lower HF QPO is different among sources
 with different luminosity, i.e. accretion rate. The amplitude of the 
lower HF QPO is up to $\sim20\%$ in atoll sources, while in Z sources 
is $\sim5\%$. Z sources have a luminosity ($\sim10^{38}$ erg/s) that can be up
 to two orders of magnitude higher than atoll ($\sim10^{36}$ erg/s) 
\cite{2006MNRAS.371.1925M}. The unique source \mbox{XTE J1701-462} displays
 both phases, i.e. atoll and Z \cite{2010MNRAS.408..622S}. In fig.~4 of 
Ref.~\cite{2010MNRAS.408..622S} is shown the amplitude (in percent of the total
 luminosity of the source) in the atoll and Z phase. The same source displays
 different amplitudes of the lower HF QPO ($10\%$ in the atoll phase, $3\%$ in
 the Z phase) at the same central frequency of the peak. Thus, it has been
 suggested that the amplitude of the lower HF
 QPO as function of its central frequency within a source 
 (e.g. figs.~\ref{fig4}, \ref{fig5} in this paper, fig.~1 in Ref.~\cite
 {2001ApJ...561.1016M}, fig.~3 in Ref.~\cite{2006MNRAS.370.1140B} and
 fig.~2 in Ref.~\cite{2006MNRAS.371.1925M}) may not be simply used as clue
 related to a signature of ISBO \cite{2010MNRAS.408..622S}. The study proposed
 in this manuscript may offer a possible interpretation to these differences
 between atoll and Z sources.

In the Z phase XTE J1701-462 has a luminosity of $L\sim0.5 L_{Edd}\sim10^{38}$ erg/s (fig.~5
 in Ref.~\cite{2010MNRAS.408..622S}), where $L_{Edd}\sim2.5\times10^{38}$ erg/s
 is the
 Eddington luminosity for a $\sim2 M_{\odot}$ neutron star \cite{2002apa..book.....F}.
 The accretion
 rate of the source is $\dot{M}\sim7\times10^{17}$ g/s. The lower HF
 QPO has a coherence up to $Q\sim30$ and amplitude $\sim3\%$, corresponding to 
 $\sim3\times10^{36}$ erg/s. At lower accretion rates ($\dot{M}\sim7\times10^{16}$ g/s)
 in the atoll phase the coherence is up to $Q\sim150$ and the amplitude up to 
 $\sim10\%$, also corresponding to $\sim2\times10^{36}$ erg/s (for such details
 see fig.~5 in Ref.~\cite{2010MNRAS.408..622S}).
 From the calculations in Sec.~\ref{sec42} and Sec.~\ref{sec5} we obtain that in
 the Z phase structures as large as $R\sim3000$ m can survive to tides, making
 $10$ periastron passages, producing an oscillation with coherence $Q\sim20$
 (typical of the $Q$ observed in XTE J1701-462 in its Z phase 
 \cite{2010MNRAS.408..622S}) 
 and emitting the observed energy of the lower HF QPO $\sim3\times10^{36}$ erg,
 corresponding to $\sim0.6\%\ \mu c^{2}$.
 The limiting size $R_{l}$ given by the accretion rate is $R_{l}\sim900$ m. The
 initial clump with $R=3000$ m spirals over an interval of orbits of 
 $0.73\ r_{g}\sim2100$ m.   
 
In the atoll phase of XTE J1701-462 the accretion rate is 
 $\dot{M}\sim7\times10^{16}$ g/s and the lower HF QPO has also a luminosity 
 $\sim10^{36}$ erg/s ($10\%$ of the luminosity of the source). Because of the 
 lower accretion rate, the density is lower
 by a factor $\sim5$ ($\rho\sim1$ g/cm$^{3}$) than in Z phase. A
 structure with $R\sim3000$ m can spiral over $1.5\ r_{g}$ and emits 
 $\sim2\times10^{36}$ erg, corresponding to $1.3\%\ \mu c^{2}$. 
 The limiting size is $R_{l}\sim500$ m and
 this last stage lasts for $\sim60$ periastron passages, spiraling over 
 $0.2\ r_{g}$. Thus, the lower
 HF QPO has a coherence of $Q\sim130$, much higher than that in the Z phase, as
 in the observations \cite{2010MNRAS.408..622S}. 
\begin{figure}
\includegraphics[width=0.47\textwidth]{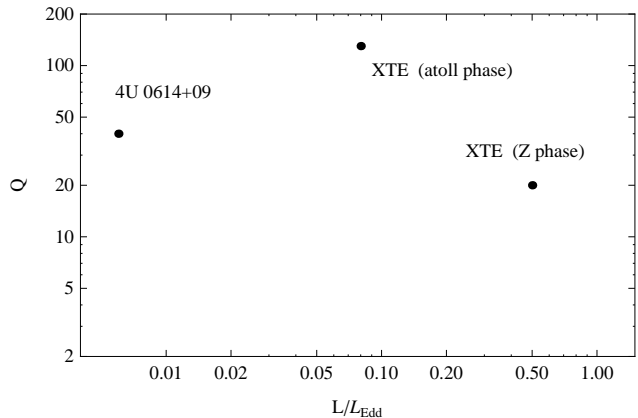}
\caption{Simulated quality factor $Q$ (coherence) of the Keplerian modulation
 by an orbiting clump of plasma as function of the luminosity $L$ of the source
 (in units of the Eddington luminosity $L_{Edd}\sim2.5\times10^{38}$ erg/s) for
 two representative sources (see text). Such coherences are typical of the
 lower HF QPO in these sources (see fig.~5 in Ref.~\cite{2010MNRAS.408..622S}
 and fig.~3 in Ref.~\cite{2006MNRAS.371.1925M} for a comparison with the data).}
\label{fig6}
\end{figure}

We did the calculations also for the source 4U 0614+09, which has the lowest 
luminosity in fig.~5 in Ref.~\cite{2010MNRAS.408..622S} (see also fig.~3 in
 Ref.~\cite{2006MNRAS.371.1925M}). Its accretion rate is 
$\dot{M}\sim7\times10^{15}$ g/s and the maximum luminosity of the lower HF
 QPO is $\sim10^{35}$ erg/s ($\sim10\%$ that of the source). The limiting size
 given by the typical speed of sound $c_{s}$ is 
 $R_{l}\sim400$ m, the density of the plasma $\rho\sim0.5$ g/cm$^{3}$ (one
 order of magnitude lower than Z sources). An initial structure with $R\sim2500$
 m spirals by $0.8\ r_{g}$, $0.4\ r_{g}$, $0.2\ r_{g}$, $0.1\ r_{g}$ and 
 $0.03\ r_{g}$ after the first, second, etc  periastron passage. Afterwards the
 relic reaches the limiting size $R\sim R_{l}$ and spirals by $0.003\ r_{g}$ per
 periastron passage making $\sim15$ passages. In total, it might make $\sim20$
 periastron passages before disrupting ($Q\sim40$). The energy emitted is 
 $\sim3\times10^{35}$ erg corresponding to $\sim1.2\%\ \mu c^{2}$.\\ 
 In Z sources the clump makes a similar number of turns before disrupting. The
 allowed number of turns by tides might be dictated by the bigger size $R_{l}$,
 allowing tides to make more work per periastron passage.
 
Fig.~\ref{fig6} shows the simulated coherence $Q$ of the Keplerian oscillation
 obtained with such mechanism for both the atoll and Z phase of XTE J1701-462
 and the lowest luminosity source 4U 0614+09. The figure reproduces the
 observed coherence of the lower HF QPO as function of the luminosity
 (accretion rate) of the source (fig.~5 in Ref.~\cite{2010MNRAS.408..622S} and
 fig.~3 in Ref.~\cite{2006MNRAS.371.1925M}). These results suggest that the
 lower HF QPO might correspond to the Keplerian modulation of orbiting matter
 around the compact object, similar to what the results in Sec.~\ref{sec5}
 suggest.

From the mechanism proposed here, we see that structures of plasma with size 
$R\sim3\times10^{3}$ m (at an orbital radius $r\sim8\ r_{g}\sim24$ km, $R/r\sim0.1$)
 are required to produce the observed luminosity of the lower HF QPO
 among a sample of sources with different accretion rates. They would emit an
 energy that is $\sim10^{36}$ erg in both atoll and Z sources ($\sim10\%$ and 
$\sim3\%$ the total luminosity, respectively) and in the lowest luminosity
 state (e.g. 4U 0614+09 with density of the plasma one order of magnitude
 lower) $\sim10^{35}$ erg. They might produce Keplerian modulations with
 different coherences similar to the observations (fig.~\ref{fig6}), because of
 the interplay between tides and the mechanical properties of the plasma.\\
 Speaking about orders of magnitude, we notice that the size of the 
 structure required to emit the observed energy is
 similar among a sample of sources spanning $\sim2$ orders of
 magnitude in luminosity. The geometry of the space-time because of the
 neutron star might play a relevant role among the sample of sources.
 Larger structures may not survive/form at all because of tides. This may
 justify why we do not observe an amplitude of the lower 
 HF QPO of, e.g., $10-20$\% in Z sources ($\sim10^{37}$ erg/s), or higher than
 those observed in atoll sources.\\ 
 From (\ref{eq9}) we derive $\sigma/Y\sim70$
 in the Z state of XTE J1701-462 and also for the Z source in 
 Sec.~\ref{sec42}. \mbox{$\sigma/Y\sim300$} in the atoll
 states, i.e. both the atoll phase of XTE J1701-462 and 4U 0614+09, even if
 they differ by one order of magnitude in luminosity. Thus, the atoll and Z
 states of LMXBs may characterize some properties of the plasma in the
 accretion disk (e.g. electrochemical bonds and/or magnetohydrodynamical).
 
To conclude, different accretion flows between Z and atoll sources might change
 the mechanical properties of the clumps of plasma, such as $\rho$, $c_{s}$, $Y$
 and $\sigma/Y$, in a way that both the amplitudes and coherences of the lower
 HF QPO also depends on the accretion flow \cite{2010MNRAS.408..622S}.
 However, within a source that does not change dramatically its accretion rate 
 the behavior of the amplitude of the lower HF QPO with its central frequency
 (i.e. figs.~\ref{fig4}, \ref{fig5} in this paper, fig.~1 in 
 Ref.~\cite{2001ApJ...561.1016M}, fig.~3 in Ref.~\cite{2006MNRAS.370.1140B} and
 fig.~2 in Ref.~\cite{2006MNRAS.371.1925M}), after this analysis, may still
 remain a candidate for a signature of ISBO.

\section{Remarks and conclusions}\label{sec8}

\citet{2012Sci...337..949R} reported a QPO in the X-ray flux from the tidal
 disruption of a star by a supermassive black hole. Here we have shown that
 the amplitude of the lower HF QPO in NS LMXBs might display the Schwarzschild/Kerr
 potential energy released by spiraling clumps of plasma in the accretion
 disk. By means of first-approximation calculations this mechanism gives
 orders of magnitude typical of the observed
 energy carried by the lower HF QPO. We would stress that magnetohydrodynamic
 modeling is certainly required to reach firmer conclusions. The  estimations
 reported here are promising and are shown for the first time.    

The mechanical properties of the plasma 
($c_{s}$, $Y$, $\sigma/Y$) in the accretion disk  
 depend on its density and temperature and, therefore, on the accretion rate
 of the source. This study suggests that both the amplitude and
 coherence of the lower HF QPO also might depend on the interplay between the mechanical properties
 of the clumps ($c_{s}$, $Y$, $\sigma/Y$) and the work done by tides on them.
 Such differences among sources with different accretion rate are reported in Ref.
\cite{2006MNRAS.371.1925M,2010MNRAS.408..622S} and were also discussed in 
Ref.~\cite{2007MNRAS.376.1139B}.    
 
The main and new results of this study are: (i)
 the differences of Schwarzschild/Kerr potential energy
 emitted by spiraling clumps of plasma might account for the observed energy
 (amplitude) carried by the lower HF QPO. Also, the
 coherence time obtained from the time-scale of tidal disruption 
 ($\sim0.03-0.2$ s) is typical of the observed ones ($\sim0.01-0.2$ s; 
 Refs.~\cite{2006MNRAS.370.1140B,2011ApJ...728....9B,2006MNRAS.371.1925M,2010MNRAS.408..622S}). This 
 is the first time that a proposed physical mechanism can justify both the observed emitted energy
 and coherence of the lower HF QPO. We are able to give an interpretation to the
 different coherence of the lower HF QPO among a sample of sources with different luminosity 
 (fig.~\ref{fig6}).  (ii) Within a source with a given accretion rate, the typical 
 bell-shape of the amplitude of
 the lower HF QPO as function of its central frequency (seen in the data collected from several 
 NS LMXBs 
 \cite{2001ApJ...561.1016M,2006MNRAS.370.1140B,2006MNRAS.371.1925M,2011ApJ...728....9B}) might be 
 reproduced by
 differences of Schwarzschild/Kerr potential energy between near orbits over the
 region $r\sim6-10\ r_{g}$.  Thus, such bell-shaped behavior may be related
 to the terms $\propto 1/r^{3}$ in (\ref{eq1}) and (\ref{eq17}) in a strong field regime and
 therefore could be a geometry-related effect due to ISBO (figs.~\ref{fig2}, \ref{fig4} and \ref{fig5}). 
 We however would stress that the Schwarzschild and Kerr metrics only
 approximate the space-time around a neutron star \cite{2003MNRAS.342..274M}.
 (iii) Our calculations might justify for the first time the interval of radius where the
 lower HF QPO is produced, i.e. $\sim6-10\ r_{g}$: In this region the highest difference of
 potential energy is released between the orbits
 the clumps spiral over before disrupting (fig.~\ref{fig2}). Future observations
 with a higher signal to noise ratio \cite{2012ExA....34..415F} could extend
 this region which, with the data to date, it may just be masked by the
 background radiation from the disk.

\begin{acknowledgments}
We thank the anonymous referee for constructive criticism that improved this manuscript.
CG thanks the program PNPD/CAPES-Brazil for full support. RC is grateful to CNPq, CAPES and FAPEMA (Brazilian agencies) by the financial support. We thank Adalto Gomes, Manoel Ferreira Jr. and  Rodolfo Angeloni for the stimulating discussions on the topic.
\end{acknowledgments}

\bibliography{biblio.bib}

\end{document}